\documentclass[preprint,aps,
showpacs
]{revtex4}

\usepackage{epsfig}

\input{epsf}

\newcommand{\be}{\begin{equation}}
\newcommand{\ee}{\end{equation}}
\newcommand{\bea}{\begin{eqnarray}}
\newcommand{\eea}{\end{eqnarray}}
\newcommand{\nn}{\nonumber \\}

\newcommand{\Bn}{{\bf n}}

\begin{document}

\preprint{Guchi-TP-014}
\date{\today%
}
\title{Multi-graviton theory, a latticized dimension, and
the cosmological constant}

\author{Nahomi Kan}
\email{b1834@sty.cc.yamaguchi-u.ac.jp}
\affiliation{Graduate School of Science and Engineering, Yamaguchi University, 
Yoshida, Yamaguchi-shi, Yamaguchi 753-8512, Japan}
\author{Kiyoshi Shiraishi}
\email{shiraish@po.cc.yamaguchi-u.ac.jp}
\affiliation{Faculty of Science, Yamaguchi University,
Yoshida, Yamaguchi-shi, Yamaguchi 753-8512, Japan}

\begin{abstract}

Beginning with the Pauli-Fierz theory, we construct a model for
multi-graviton theory. Couplings between gravitons belonging to
nearest-neighbor ``theory spaces'' lead to a discrete mass spectrum.
Our model coincides with the Kaluza-Klein theory whose fifth dimension
is latticized.

We evaluate one-loop vacuum energy in models with a circular latticized
extra dimension as well as with compact continuous dimensions. 
We find that the vacuum energy can take a positive value, if the
dimension of the continuous space time is $6, 10,\ldots$.
Moreover, since the amount of the vacuum energy can be an arbitrary small
value according to the choice of parameters in the model,
our models is useful to explain the small positive dark energy in the
present universe.

\end{abstract}

\pacs{04.50.+h, 04.60.Nc, 11.10.Kk}


\maketitle


\section{Introduction and summary}

It is well known that the weak-field limit of the Einstein gravity
reduces to the theory of a single massless spin-two field~\cite{spin2}.
We call the second-order symmetric tensor field $h_{\mu\nu}$,
which represents deviation from a flat space
($g_{\mu\nu}=\eta_{\mu\nu}+h_{\mu\nu}$),  as a ``graviton''.
Multi-graviton theories have been studied by
Boulanger {\it et al.}~\cite{Boulanger,BDGH}.
They have shown that there is no consistent interaction of gravitons
and the possible action is a sum of replicated Pauli-Fierz
actions~\cite{PF}.

In this paper, we consider a model for a multi-graviton theory with
nearest-neighbor couplings in the theory space.
To be precise, the gravitons are not interacting each other in our model,
but they have a discrete mass spectrum.
The simplest model is explained in Sec.~\ref{sec:2}.
In the limit of a large number of gravitons,
the mass spectrum of gravitons resembles 
that of the Kaluza-Klein (KK) theory. 
Therefore our model is equivalent to the KK theory
with a discretized dimension.

The emergence mechanisms of a space dimension have been suggested in
refs.~\cite{Sugamoto,Bander,JLM}, which is in the stream of dimensional
deconstruction~\cite{ACG,HPW}. Recently the authors have been
informed of ref.~\cite{AGS}, in which the gravity with the theory space
is also argued. Although our model for the gravitation theory with a
discretized dimension is very simple, it may be a toy model 
adequate for studying qualitative features of more complicated theory
and mechanisms included in it.

For example, we evaluate the vacuum energy in our model in
Sec.~\ref{sec:3}.
We also consider a model in the continuous $(4+\delta)$ dimensional
space-time with a discretized dimension.
We show that an arbitrary small amount of positive vacuum energy can be
obtained if $(4+\delta)=6, 10, \ldots$.

\section{multi-graviton theory and a latticized dimension}
\label{sec:2}

The lagrangian for a massless spin-two field can be written as
\be
{\cal L}_0=-\frac{1}{2}\partial_{\lambda}h_{\mu\nu}
\partial^{\lambda}h^{\mu\nu}+\partial_{\lambda}h^{\lambda}{}_{\mu}
\partial_{\nu}h^{\nu\mu}-\partial_{\mu}h^{\mu\nu}
\partial_{\nu}h+\frac{1}{2}\partial_{\lambda}h
\partial^{\lambda}h\, ,
\ee
where $h\equiv h^{\mu}_{\mu}$.
This corresponds with the weak-field limit of the Einstein gravity.

The lagrangian for a massive graviton with the St\"uckerberg
fields~\cite{St} is known as~\cite{Hamamoto}
\be
{\cal L}_m={\cal L}_0-\frac{m^2}{2}(h_{\mu\nu}h^{\mu\nu}-h^2)
-2(mA^{\mu}+\partial^{\mu}\varphi)
(\partial^{\nu}h_{\mu\nu}-\partial_{\mu}h)-\frac{1}{2}
(\partial_{\mu}A_{\nu}-\partial_{\nu}A_{\mu})^2\, ,
\label{mpf}
\ee
where a constant $m$ corresponds to the mass of graviton.
We have omitted here the gauge fixing term as well as the ghost terms.
The lagrangian (\ref{mpf}) is invariant under the following
transformations:
\bea
h_{\mu\nu}&\rightarrow&h_{\mu\nu}+\partial_{\mu}
\xi_{\nu}+\partial_{\nu}\xi_{\mu}\, ,\\
A_{\mu}&\rightarrow&A_{\mu}+m\xi_{\mu}-\partial_{\mu}\zeta\, ,\\
\varphi&\rightarrow&\varphi+m\zeta\, ,
\eea
where $\xi$ and $\zeta$ are local parameters. 

Next we consider a multi-graviton
($N$-graviton) theory~\cite{Boulanger,BDGH} given by the lagrangian
\be
\bar{\cal L}_0=\sum_{k=1}^N
\left[-\frac{1}{2}\partial_{\lambda}h_{\mu\nu}^k
\partial^{\lambda}h^{k\,\mu\nu}+\partial_{\lambda}h^{k\,\lambda}{}_{\mu}
\partial_{\nu}h^{k\,\nu\mu}-\partial_{\mu}h^{k\,\mu\nu}
\partial_{\nu}h^k+\frac{1}{2}\partial_{\lambda}h^k
\partial^{\lambda}h^k\right]\, .
\label{ngra}
\ee
It can be said that a massless graviton lives in each ``theory space'',
which is labeled by $k$.

Now, instead of adding the mere mass term to the lagrangian (\ref{ngra}),
we put the
term including couplings of other gravitons with the nearest-neighbor
suffix:
\bea
\bar{\cal L}_m&=&\bar{\cal L}_0
-\frac{m^2}{2}\sum_{k=1}^N\left[
(h^k_{\mu\nu}-h^{k+1}_{\mu\nu})^2-(h^k-h^{k+1})^2\right]\nn
& &-2\sum_{k=1}^N\left[m(A^{k\,\mu}-A^{k-1\,\mu})+
\partial^{\mu}\varphi^k\right]
(\partial^{\nu}h^k_{\mu\nu}-\partial_{\mu}h^k)
-\frac{1}{2}\sum_{k=1}^N
(\partial_{\mu}A^k_{\nu}-\partial_{\nu}A^k_{\mu})^2\, .
\label{dis}
\eea
In this notation, one should read as $h^{k+N}=h^k$, $h^{0}=h^N$,
$h^{N+1}=h^1$, and so on.

It is interesting to see that the multi-graviton lagrangian
is invariant under the transformations: 
\bea
h^k_{\mu\nu}&\rightarrow&h^k_{\mu\nu}+\partial_{\mu}
\xi^k_{\nu}+\partial_{\nu}\xi^k_{\mu}\, ,\nn
A^k_{\mu}&\rightarrow&A^k_{\mu}+m\xi^k_{\mu}-m\xi^{k+1}_{\mu}
-\partial_{\mu}\zeta^k\, ,\nn
\varphi^k&\rightarrow&\varphi^k+m\zeta^k-m\zeta^{k-1}\, .
\label{trans}
\eea

As in the case with 5D QED studied by Hill and Leibovich~\cite{HL1},
$N-1$ massive gravitons and one massless graviton are described
by the lagrangian.
The eigenvalues for the mass matrix are 
\be
M_{p}^2=4m^2\sin^2\left(\frac{\pi p}{N}\right)\qquad (p=1,\ldots, N)\, .
\ee
Massive vector and scalar fields are absorbed by the massive graviton
fields by means of the transformations (\ref{trans}) and then they have
correct spin-degree of freedom, five.

On the other hand, massless vector and scalar fields are survived.
The massless degree of freedom of the spin-two, one, zero fields are
described as
\be
\bar{h}_{\mu\nu}=\frac{1}{\sqrt{N}}\sum_{k=1}^N h^k_{\mu\nu}\, ,\qquad
\bar{A}_{\mu}=\frac{1}{\sqrt{N}}\sum_{k=1}^N A_{\mu}^k\, ,\qquad
\bar{\varphi}=\frac{1}{\sqrt{N}}\sum_{k=1}^N \varphi^k\, .
\ee 
When we make the following combination
\be
H_{\mu\nu}=\bar{h}_{\mu\nu}+\bar{\varphi} \eta_{\mu\nu}\, ,
\ee 
it becomes a massless graviton and then the kinetic term of 
$\bar{\varphi}$ takes a canonical form~\cite{Hamamoto}.

We find that the field content of our model is very akin to
the one of the KK theory~\cite{KK}. Suppose that we considered
the five-dimensional metric as
\be
ds^2=(\eta_{\mu\nu}+h_{\mu\nu})dx^{\mu}dx^{\nu}-2A_{\mu}dx^{\mu}dy+
(1+2\varphi)dy^2\, ,
\ee
or
\be
A_{\mu}=h_{\mu y}~\, ,\qquad \varphi=h_{yy}~\, .
\ee
When we take the weak-field limit of the five-dimensional
Einstein-Hilbert action, the lagrangian density appears to be of the
form~(\ref{dis}),
provided that the derivative with respect to the extra dimensional
coordinate $y$ is replaced by the difference operation, namely
\be
\frac{\partial h_{\mu\nu}}{\partial y}\longrightarrow
\Delta_y h_{\mu\nu}\equiv
\frac{h_{\mu\nu}(x,y+\epsilon)-h_{\mu\nu}(x,y)}{\epsilon}\, ,
\ee
and we define
$h_{\mu\nu}^k(x)\equiv h_{\mu\nu}(x,k\epsilon)$ with $\epsilon=m^{-1}$.

The mass spectrum and spin degrees of freedom in the
KK theory have been explained by ref.~\cite{CZ}.
Our multi-graviton model can be regarded as a discrete version of the
KK theory.
Thus the
transformation~(\ref{trans}) can also be derived in the similar manner to
ref.~\cite{CZ}.

The simplest way to get actions for multi-graviton system
is the discretization of extra dimensions in generalized
KK theories.
The counting of the four-dimensional fields can be done
by the completely parallel way to the analysis in ref.~\cite{HLZ,GRW}.
If we consider the discretized version of the  KK theory
in $M_4\otimes T^d$ (where $M_4$ denotes the four-dimensional Minkowsky
spacetime and $T^d$ is a $d$ dimensional torus), 
we obtain a tower of massive
spin-two states,
$(d-1)$ tower of massive spin-one states, and
$d(d-1)/2$ tower of massive spinless states.
As well, we obtain the following
massless states:
one spin-two, $d$ spin-one, and $d(d+1)/2$ spinless states.

It is an important problem whether all of the field contents are
necessary or not in the view of the four-dimensional gravity.
In other words, we are interested in the possibility of the
multi-graviton theory whose continuum limit does not coincide with a
higher-dimensional theory.
Particularly, the presence of the massless vector and scalar fields
are unwanted when we wish to consider a phenomenological model.
Attempts to solve this problem by imposing discrete symmetries
in the theory space and
discretization of non-abelian KK theories will be considered elsewhere.

\section{vacuum energy}
\label{sec:3}

In this section, we evaluate the one-loop vacuum energy density
in the model of the previous section. We will also offer an extension of
the model, which includes both latticized and continuous extra
dimensions, and the vacuum energy is estimated in the model.

Recently, the scenarios explaining the small positive cosmological
constant in the present universe ($10^{-47} {\rm GeV}^4$, 
according to~\cite{Sahni})
 are discussed through the study
of the Casimir-like energy in the higher-dimensional 
theories~\cite{Gupta,Milton}. Because the latticized model would have
different UV behavior from that of continuum model,
we expect a novel expression of the cosmological constant.

First, we consider one scalar degree of freedom 
which has the mass spectrum discussed in the previous section.
Then the vacuum energy density  in the
one-loop calculation is written
by the zeta-function technique~\cite{Hawking} as
\be
\frac{1}{2}\ln\det[-\nabla^2+M_p^2]=-\frac{1}{2}
\left.\frac{d\zeta(s)}{ds}\right|_{s=0}\, ,
\ee
where
\be
\zeta(s)=\frac{\mu^{2s}}{16\pi^2\Gamma(s)}
\sum_p\int_{0}^{\infty}dt~t^{s-3}\exp(-M^2_p t)\, ,
\ee
with
\be
M_{p}^2=4m^2\sin^2\left(\frac{\pi p}{N}\right)\, .
\ee
Here the constant $\mu$ has the dimension of mass.

The calculation of the zeta function can be done by the similar method
used in refs.~\cite{fw,mi}.
Using the formula
\be
\exp\left[-4m^2\sin^2(\theta/2)t\right]=e^{-2m^2t}\sum_{\ell=-\infty}^{\infty}
\cos \ell\theta~ I_{\ell}(2m^2t)\, ,
\ee
where $I_{\nu}(x)$ is the modified Bessel function,
we can write the zeta function as
\be
\zeta(s)=\frac{\mu^{2s}}{(4\pi)^{2}\Gamma(s)}\sum_{p}\sum_{\ell=-\infty}^{\infty}
\cos {\frac{2\pi p\ell}{N}}~\int_0^{\infty}dt~t^{s-3}
e^{-2m^2t} I_{\ell}(2m^2t)\, .
\ee
For simplicity, we assume $N\ge 3$.
Carrying out the summation over $p$ first,
we find that only the terms of $\ell=qN~(q: integer)$ survive.
Then we find
\be
\zeta(s)=\frac{\mu^{2s}N}{(4\pi)^{2}\Gamma(s)}\left[
\int_0^{\infty}dt~t^{s-3}
e^{-2m^2t} I_{0}(2m^2t)+2
\sum_{q=1}^{\infty}
~\int_0^{\infty}dt~t^{s-3}
e^{-2m^2t} I_{qN}(2m^2t)\right]\, .
\ee

According to ref.~\cite{GR}, the integration can be carried out
and is found to be
\be
\int_0^{\infty}dt~t^{s-3}
e^{-2m^2t} I_{\ell}(2m^2t)=(4m^2)^{2-s}
\frac{\Gamma(\frac{5}{2}-s)\Gamma(\ell-2+s)}%
{\sqrt{\pi}\Gamma(\ell+3-s)}\, .
\ee
Therefore the zeta function is written as
\be
\zeta(s)=\frac{N\,
m^4}{\pi^{2}}\left(\frac{\mu^2}{4m^2}\right)^s
\frac{\Gamma(\frac{5}{2}-s)}{\sqrt{\pi}}\left[
\frac{1}{(s-1)(s-2)\Gamma(3-s)}+\frac{2}{\Gamma(s)}
\sum_{q=1}^{\infty}
~\frac{\Gamma(qN-2+s)}%
{\Gamma(qN+3-s)}\right]\, .
\ee
Now we obtain
\be
-\left.\frac{d\zeta(s)}{ds}\right|_{s=0}=-\frac{3N\,
m^4}{16\pi^{2}}\ln\left(\frac{\bar{\mu}^2}{4m^2}\right)
-\frac{3m^{4}}{2\pi^{2}}
\sum_{q=1}^{\infty}
\frac{1}%
{q(q^2N^2-1)(q^2N^2-4)}\, ,
\ee
where
\be
\ln\left(\frac{\bar{\mu}^2}{4m^2}\right)=
\ln\left(\frac{\mu^2}{4m^2}\right)+\psi(3)-\psi(5/2)+\frac{3}{2}\, .
\ee

Because there are five states in each mass level,
the vacuum energy in the multi-graviton model in the previous section
is
\be
V=V_R-\frac{15N\,
m^4}{32\pi^{2}}\ln\left(\frac{\bar{\mu}^2}{4m^2}\right)
-\frac{15m^{4}}{4\pi^{2}}
\sum_{q=1}^{\infty}
\frac{1}%
{q(q^2N^2-1)(q^2N^2-4)}\, ,
\ee
where $V_R(\bar{\mu})$ is the renormalized vacuum energy.

According to ref.~\cite{Cherednikov}, we require the invariance of $V$
with respect to the scale $\bar{\mu}$ and we choose $V_R(\bar{\mu}_R)=0$
for a normalization point $\bar{\mu}_R$. Then we find
\be
V=-\frac{15N\,
m^4}{32\pi^{2}}\ln\left(\frac{\bar{\mu}_R^2}{4m^2}\right)
-\frac{15m^{4}}{4\pi^{2}}
\sum_{q=1}^{\infty}
\frac{1}%
{q(q^2N^2-1)(q^2N^2-4)}\, .
\ee

Unfortunately, this vacuum energy density is negative,
for $\bar{\mu}_R$ is considered as a cutoff due to a scale of a new
physics, $\bar{\mu}_R>m$. The boundary condition $V_R(\bar{\mu}_R)=0$
can be interpreted as the expectation that the new physics forces the
vacuum energy or cosmological constant to vanish at such a high-energy
scale.

Let us consider the extension of our model to the one with
$4+\delta$ dimensional continuous spacetime.
Here the extra $\delta$ dimensional space is taken as 
$(S^1)^{\delta}$ with a common radius $b$.
One can regard the new model as the theory with a latticized circle
and $\delta$ continuous circles.
In this model the mass spectrum of the gravitons is expressed as
\be
M_{p,\Bn}^2=4m^2\sin^2\left(\frac{\pi p}{N}\right)+
\frac{\Bn^2}{b^2}\, ,
\ee
where
$\Bn=(n_1, n_2,\ldots, n_{\delta})$
and all the components take integer values.

Using the Poisson resummation formula
\be
\sum_{\Bn}\exp\left(-\frac{\Bn^2}{b^2}t\right)=
\left(\frac{\pi b^2}{t}\right)^{\delta/2}
\sum_{\Bn}\exp\left(-\frac{\pi^2b^2\Bn^2}{t}\right)\, ,
\ee
we can express the zeta function, in which in the present time
the summation over $\Bn$ is involved, as
\bea
\zeta(s)&=&\frac{\mu^{2s}N(2\pi
b)^{\delta}}{(4\pi)^{2+\delta/2}\Gamma(s)}\sum_{\Bn}\left[
\int_0^{\infty}dt~t^{s-\delta/2-3}
\exp\left(-2m^2t-\pi^2b^2\Bn^2/t\right) I_{0}(2m^2t)\right.\nn
& &\left.+2
\sum_{q=1}^{\infty}
~\int_0^{\infty}dt~t^{s-\delta/2-3}
\exp\left(-2m^2t-\pi^2b^2\Bn^2/t\right) I_{qN}(2m^2t)\right]\, .
\label{ff}
\eea
If $\delta$ is odd, all the terms in the parentheses are finite
and lead to no logarithm term in the vacuum energy. The vacuum energy
can be found to be finite and negative in this case, as in the case of
the KK theory with torus compactification.
Thus we assume that $\delta$ is even. 

For even $\delta$, in the limit $s\rightarrow 0$,
only  the first term with $\Bn={\bf 0}$ is divergent in the parentheses
on the right hand side of (\ref{ff}) if $N$ is larger than $2+\delta/2$.
Thus, after taking account of the degrees of freedom, the vacuum energy
density in our extended model takes the form
\be
V=V_R-(-1)^{\delta/2}\frac{(5+\delta)(2+\delta)N\,
m^4}{4\pi^{2}}(4\pi m^2b^2)^{\delta/2}
\frac{\Gamma(\frac{5+\delta}{2})}{\sqrt{\pi}\Gamma(3+\frac{\delta}{2})^2}
\ln\left(\frac{\bar{\mu}^2}{4m^2}\right)
-\cdots\, ,
\ee
where
\be
\ln\left(\frac{\bar{\mu}^2}{4m^2}\right)=
\ln\left(\frac{\mu^2}{4m^2}\right)+2\psi(3+\delta/2)-\psi(5/2+\delta/2)+
\gamma\, ,
\ee
($\gamma$ is the Euler's constant) and $\ldots$ means the positive
finite terms.

As seen from the sign of logarithmic term, we can obtain the positive
vacuum energy after taking account of the renormalization
discussed above; in
other words, we have the positive cosmological constant when 
$\delta=4 k+2$ $(k=0,1,2,\ldots)$,
such as $\delta=2$
($D=4+\delta=6$), $\delta=6$ ($D=10$), and so on.

We will further investigate the possibility of the arbitrarily small
cosmological constant. Since the contribution from the finite $q$ terms
in the zeta function can be estimated as at most $O(N^{-4})$, it can be
neglected if a large $N$ is assumed. Further if $mb\ll1$, we find
\bea
V&\approx&\frac{(5+\delta)(2+\delta)N\,
m^4}{4\pi^{2}}(4\pi m^2b^2)^{\delta/2}
\frac{\Gamma(\frac{5+\delta}{2})}{\sqrt{\pi}\Gamma(3+\frac{\delta}{2})^2}
\ln\left(\frac{\bar{\mu}_R^2}{4m^2}\right)\nn
&-&\frac{(5+\delta)(2+\delta)N}{4\pi^{(4+\delta)/2}(2\pi
b)^{4}}\Gamma\left(\frac{4+\delta}{2}\right)
\sum_{\Bn\ne{\bf 0}}\frac{1}{(\Bn^2)^{(4+\delta)/2}}\, 
\qquad (mb\ll 1).
\eea
Therefore when $(2\pi mb)^{4+\delta}\ln(\bar{\mu}_R/m)\approx O(1)$,
an arbitrary small vacuum energy can be obtained.
Note that the condition does not include $N$ in this approximation.

The difficulty of the present model in this section is
the problem of tuning the scale of the extra dimensions.
The model does not provide a mechanism to determine the value of $b$,
because there is no minimum value of $V$ as a function of $b$. 
We should consider either the value of $b$ is due to the other mechanism
or $b$ takes an appropriate value only in the present universe.
For the latter case, the motion of $b$ may play a role of quintessence or
dark energy, as similar to the suggestion in ref.~\cite{WR}. This
possibility will be considered elsewhere.

In this paper, we only considered the vacuum energy at the one-loop
level. Even at the same level, the correction to the Newton constant
may also be induced in general. This would be related with the
restriction or determination of the scale of the extra dimensions as
well as the scale of $m$. 

We should investigate the nature of coupling to the matter fields
and the back reaction problem in the case with a small cosmological
constant. Through such studies, the case with strong
gravity will also be revealed.

Finally, we remember that we have not treated the hierarchy problem
in this paper.
It will be clarified by the further study whether incorporating
supersymmetry (and its break-down) into the model is necessary or
not.


\begin{acknowledgments}
The authors thank the Yukawa Institute for Theoretical Physics at 
Kyoto University. Discussions during the YITP workshop YITP-W-02-18 on
``Quantum Field Theories: Fundamental Problems and Applications'' were
useful to complete this work. 

We would like to thank K.~Sakamoto and Y.~Cho for his valuable comments
and for the reading the manuscript.
\end{acknowledgments}



\end{document}